\pdfoutput=1
\documentclass[journal=mamobx,manuscript=article,layout=twocolumn,email=true]{achemso}

\usepackage[T1]{fontenc} 
\usepackage{graphicx}
\usepackage[dvipsnames]{xcolor}
\usepackage{amsmath}
\usepackage{amssymb}
\usepackage{multirow}
\usepackage{mathtools}
\usepackage{hyperref}
\hypersetup{
    colorlinks=true,
    linkcolor=red,
    citecolor=blue,
    filecolor=green,      
    urlcolor=black
}
\usepackage{multirow}
\usepackage{multicol}
\usepackage{caption}
\makeatletter
\let\l@addto@macro\relax
\makeatother
\usepackage[fontsize=10pt]{scrextend}
\let\oldmaketitle\maketitle
\let\maketitle\relax
\author{Adri Escañuela-Copado}
\affiliation{Grupo de Física de Fluidos y Biocoloides, Departamento de Física Aplicada, Universidad de Granada, 18071 Granada, Spain}
\author{José López-Molina}
\affiliation{Grupo de Física de Fluidos y Biocoloides, Departamento de Física Aplicada, Universidad de Granada, 18071 Granada, Spain}
\author{Matej Kandu\v{c}}
\affiliation{Jožef Stefan Institute, SI-1000 Ljubljana, Slovenia}
\author{Ana B. Jódar-Reyes}
\affiliation{Grupo de Física de Fluidos y Biocoloides, Departamento de Física Aplicada, Universidad de Granada, 18071 Granada, Spain}
\author{\linebreak María Tirado-Miranda}
\affiliation{Grupo de Física de Fluidos y Biocoloides, Departamento de Física Aplicada, Universidad de Granada, 18071 Granada, Spain}
\author{Delfi Bastos-González}
\affiliation{Grupo de Física de Fluidos y Biocoloides, Departamento de Física Aplicada, Universidad de Granada, 18071 Granada, Spain}
\author{José M. Peula-García}
\affiliation{Departamento of Física Aplicada II, Facultad de Ciencias, Universidad of Málaga, 29071 Málaga, Spain}
\author{\linebreak Irene Adroher-Benítez}
\affiliation{Grupo de Física de Fluidos y Biocoloides, Departamento de Física Aplicada, Universidad de Granada, 18071 Granada, Spain}
\email{iadroher@ugr.es}
\author{Arturo Moncho-Jordá}
\affiliation{Grupo de Física de Fluidos y Biocoloides, Departamento de Física Aplicada, Universidad de Granada, 18071 Granada, Spain}
\email{moncho@ugr.es}
\alsoaffiliation{Instituto Carlos I de Física Teórica y Computacional, Facultad
de Ciencias, Universidad de Granada, 18071 Granada,
Spain}

\title{Diffusion and interaction effects on molecular release kinetics from collapsed microgels}


\begin{document}


\twocolumn[
\begin{@twocolumnfalse}
\oldmaketitle
\begin{abstract}
The transport of biomolecules, drugs, or reactants encapsulated inside stimuli-responsive polymer networks in aqueous media is fundamental for many material and environmental science applications, including drug delivery, biosensing, catalysis, nanofiltration, water purification, and desalination. The transport is particularly complex in dense polymer media, such as collapsed hydrogels, where the molecules strongly interact with the polymer network and diffuse via a hopping mechanism. In this study, we employ Dynamical Density Functional Theory (DDFT) to investigate the non-equilibrium release kinetics of non-ionic subnanometer-sized molecules initially uploaded inside collapsed microgel particles. The theory is consistent with previous molecular dynamics simulations of collapsed poly($N$-isopropylacrylamide) (PNIPAM) polymer matrices, accommodating molecules of varying shapes and sizes. We found that, despite the intricate physico-chemical properties involved in the released process, the kinetics is predominantly dictated by two material parameters: the diffusion coefficient of the molecules inside the microgel ($D^*$) and the interaction free energy of the molecules with the microgel ($\Delta G$). Our results reveal two distinct limiting regimes: For large, slowly diffusing molecules weakly attracted to the polymer network, the release is primarily driven by diffusion, with a release time that scales as $\tau_{1/2} \sim 1/D^*$. Conversely, for small molecules strongly attracted to the polymer network, the release time is dominated by the interaction, scaling as $\tau_{1/2} \sim \exp(-\Delta G/k_{\textrm{B}} T)$. Our DDFT calculations are directly compared with an analytical equation for the half-release time, demonstrating excellent quantitative agreement. This equation represents a valuable tool for predicting release kinetics of non-ionic molecules from collapsed microgels.
\vspace{2em}
\end{abstract}
\end{@twocolumnfalse}
]


\section{Introduction}\label{sec:introduction}
Microgels, cross-linked polymer networks of sizes ranging from hundreds of nanometers to around a micrometer, remain an active area of research in disciplines like nanotechnology\cite{Lu2009,Wu2012}, materials science\cite{Zhang2010}, environmental sciences\cite{Nykaenen2007,Parasuraman2011,Chen2014,Hyk2018}, and biomedical sciences\cite{Peppas1997,Klouda2008,Stuart2010,Banerjee2015}. These colloidal particles are recognized for their stimuli-responsive behavior, which is characterized by reversible volume phase transitions in response to environmental stimuli such as changes in temperature\cite{Ding2020}, pH\cite{Kocak2017}, ionic strength\cite{Wang2016}, or solvent properties\cite{Bertolla2017}. This unique adaptability and functional versatility enable microgels to serve in a wide range of applications, from surface coating\cite{Chen2016} to optoelectronic switches\cite{Dokkhan2018}. In particular, their responsiveness to environmental changes further broadens their utility in medicine. They can act as effective containers for various molecules, such as proteins, polysaccharides, enzymes, nucleic acids and drugs\cite{Zelikin2008,Levy2008,Masoud2012,Kozlovskaya2014, McMasters2017,Qiao2020}, providing a foundation for advancements in drug accumulation, release, and targeted delivery\cite{Kabanov2009,Oh2008,Guan2011}. Such capabilities underscore the significant promise of microgels in driving the evolution of drug delivery systems and further developments in nanotechnology and nanomedicine.\cite{Vinogradov2010,Kureha2018}

Poly(N-isopropylacrylamide) (PNIPAM) is one of the most extensively studied thermoresponsive polymers, mainly because it exhibits a volume transition in water from a swollen to collapsed state at temperatures closely approximating the human body temperature~\cite{Pelton2000,Gil2004,Halperin2015}. Given its versatility, PNIPAM microgels have served as fundamental model systems, paving the way for numerous advancements in the field of soft responsive materials~\cite{Lu2006,Ballauff2007}.

The control over the uptake and release kinetics of molecules from nanoparticulate systems is crucial for realizing their full potential in a range of applications. For targeted drug delivery, the precise modulation of the kinetics is key to achieving desired therapeutic outcomes\cite{Hoffman1987,Gehrke1997}. Properly tuned release rates can ensure consistent drug concentrations in the systemic circulation, while efficient uptake directly impacts drug loading capacity, influencing overall therapeutic efficacy. In industrial contexts, control over uptake and release kinetics can directly influence process efficiency in catalysis or chemical separation, where the kinetics can dictate reaction rates and separation performance\cite{Horecha2014,Herves2012}. When utilized as nanoreactors, the responsive network structure of microgel particles allows for modulation of the permeability of reactants, thereby determining the reaction rate~\cite{Lu2006,Roa2017}. Therefore, understanding and manipulating these kinetics is a central aspect of nanocarrier optimization, providing a solid foundation for more predictable and reliable system performance across various applications. This understanding is essential not only for the practical implementation of these systems but also for the development of accurate theoretical models.

In most of the above-mentioned applications, the release process takes place when the microgel nanoparticle is in a collapsed state. In this conformation, the interaction of the molecule with the polymer matrix and its transport through such a crowded environment becomes significantly complex due to polymer-water interactions and obstruction effects~\cite{Merrill1993}. On the one hand, the partition ratio (concentration of molecules inside the microgel over that in bulk)  observed in the experiments is usually several times larger than the ones predicted by simple size-exclusion theories. The partitioning ratio for hydrophobic molecules (typically drugs) can even reach several orders of magnitude~\cite{Palasis1992,Guilherme2003,Molina2012}. These results have also been confirmed in atomistic computer simulations, for which the calculated transfer free energies from the bulk solution to the interior of the collapsed polymer gel can reach considerable negative values~\cite{Kanduc2018,Kanduc2021,Moncho-Jorda2020}. On the other hand, the diffusive transport of the cargo molecules across the collapsed polymer network is also very complicated~\cite{Muller1998} and is intrinsically different from diffusion in liquids and dilute systems~\cite{Takeuchi1990,Mueller‐Plathe1991}. Indeed, molecular dynamics simulations point out that the distribution of water in dense polymers highly depends on the polarity of the polymers: water molecules can be homogeneously distributed~\cite{Karlsson2004} or form clusters~\cite{Fukuda1998}. For the case of hydrophobic PNIPAM collapsed networks, experiments~\cite{Sasaki1999} and atomistic  simulations~\cite{kanduc2019free,Kanduc2021} indicate that water molecules form disconnected clusters. These clusters have a fractal-like structure, sorbed between the voids made by the polymeric chains. The diffusion of the molecules in such an obstructed network occurs via the hopping mechanism~\cite{Mueller‐Plathe1991,Muller-Plathe1992,Muller1998,Fritz1997}, in which a molecule is constrained for a longer time inside a local microscopic cavity and suddenly performs a longer jump from a local cavity to a neighboring cavity through a short-time transient water created by the thermal fluctuations of the polymer chains~\cite{Hofmann2000,Vegt2000,Neyertz2010,Kanduc2018,kanduc2019free,Kanduc2021}.

In this work, we make use of Dynamical Density Functional Theory (DDFT) to model the release of molecules previously encapsulated inside a collapsed microgel~\cite{Marconi1999,Archer2004,Wu2007,Vrugt_2023}. DDFT is a theoretical framework that has evolved as an effective tool for studying the dynamics of many-body systems in the field of condensed matter physics and soft matter science. This technique extends classical Density Functional Theory (DFT) to incorporate time-dependent phenomena, thereby enabling the investigation of out-of-equilibrium states~\cite{Hansen2013,Evans1992,Ebner1976,Ebner1977}. By allowing for the examination of time-dependent diffusive processes, DDFT provides a detailed and sophisticated approach to track cargo release.

Specifically, we construct a DDFT framework to detail the spatio-temporal evolution of the density profile of molecules encapsulated inside a collapsed microgel during the release process. This method takes into account the actual sterically-obstructed diffusion coefficient of the molecules inside the microgel. Furthermore, it incorporates the effects of both the microgel-molecule and molecule-molecule interactions. DDFT has been successfully applied to similar problems, namely the prediction of protein adsorption into nanoparticles~\cite{Angioletti-Uberti2014,Angioletti-Uberti2017,Angioletti-Uberti2018} and the encapsulation/release kinetics of neutral and charged molecules in hollow microgel particles~\cite{Moncho-Jorda2019, Moncho-Jorda2020,PerezRamirez2022}. Here, we adapt our existing DDFT framework to be consistent with atomistic simulation data from prior works~\cite{Kanduc2018,kanduc2019free,Kanduc2021}. These simulations provided values for the diffusion coefficient, $D^*$, and the transfer free energy, $\Delta G$, of a test molecule within collapsed PNIPAM polymer networks. Our results clearly indicate that these two parameters completely control the release kinetics in those systems. From the  calculations performed under many different conditions, we deduce an analytical expression for the half-release time, $\tau_{1/2}$, and we show that the time evolution of the fraction of released molecules can be scaled into a master curve that only depends on $\tau_{1/2}$.

This paper is organized as follows. After this introduction, we describe our \nameref{sec:methods}, detailing the model for cargo release from a collapsed microgel and explaining the theoretical framework applied in our study. In the following \nameref{sec:results} section we present our findings, exploring the release kinetics of molecular cargo from collapsed microgels solving the DDFT equation. We analyze the time-dependent density profile of cargo molecules as they diffuse through the polymer network, and discuss the time evolution of the fraction of released molecules. This is followed by a comprehensive study of the influence of the diffusion coefficient and the molecule--microgel interaction free energy on release dynamics. We also analyze the role of microgel size in the release behavior. One of the main contributions of this study is to show that, solving the stochastic differential equation for mean-first passage time, allows the full set of results to be gathered into a single analytical expression that correctly describes the release kinetics for any particular situation. In the \nameref{sec:conclusions} section, we summarize our primary findings and discuss their broader implications.

\section{Theoretical methods}\label{sec:methods}
\subsection{Model for cargo release from a collapsed microgel}
\begin{figure}[ht!]
    \includegraphics[width=\linewidth]{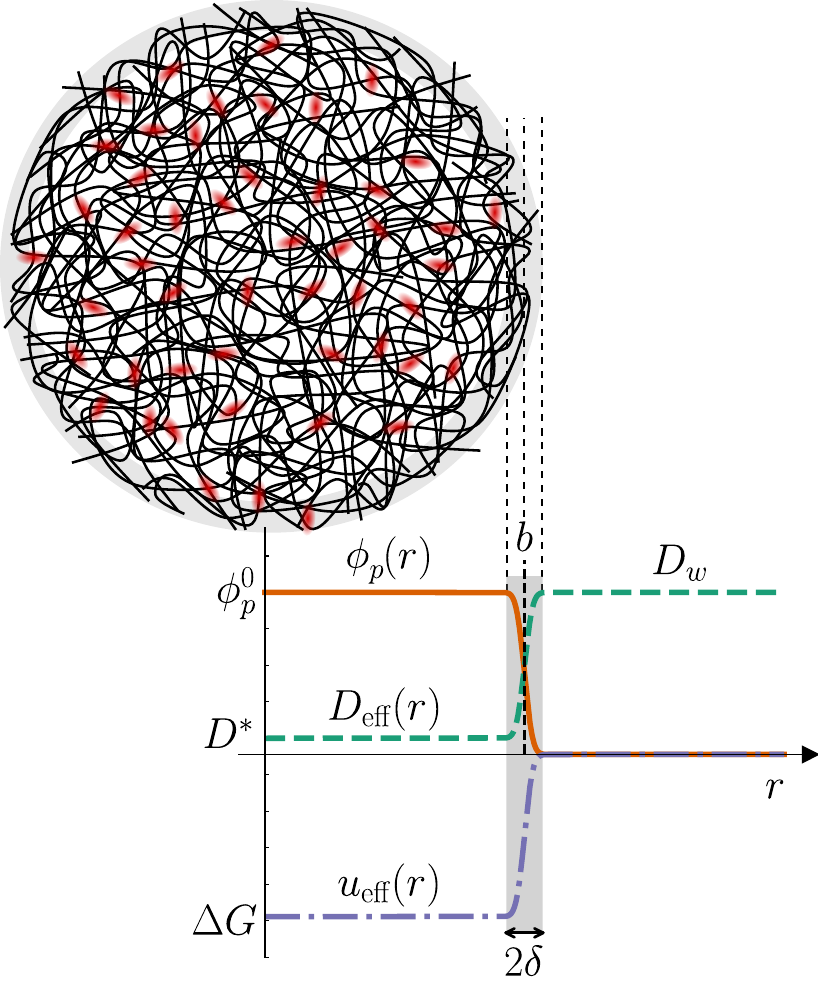}
    \caption{Representation of a collapsed microgel of radius $b$ with a narrow interface of width $2\delta\ll b$ carrying cargo molecules. The plot illustrates the radial dependence of the polymer volume fraction ($\phi_{\text{p}}(r)$), the effective diffusion coefficient of the cargo molecule ($D_{\mathrm{eff}}(r)$), which switches from $D^{*}$ inside the microgel to $D_w$ in bulk, and the effective microgel--molecule interaction ($u_{\mathrm{eff}}(r)$), which is given by the $\Delta G$ inside the microgel and decays to $0$ at the interface with the bulk.}
    \label{fig:microgel-diagram}
\end{figure}
To theoretically investigate the release of molecular cargo from a collapsed microgel, we model our system by considering a single spherical microgel particle of radius $b$ immersed in an aqueous solution, treated as a uniform background. Fig.~\ref{fig:microgel-diagram} shows a schematic illustration of the microgel particle. Although the interior of the particle exhibits localized fluctuations in polymer density, it is still possible to define an average local polymer volume fraction, $\phi_{\text{p}}$. For a collapsed microgel, $\phi_{\text{p}}$ is roughly constant inside the microgel, and abruptly declines to zero at the external interface. This radial dependence can be approximated as $\phi_{\text{p}}(r)=\phi_{\text{p}}^{0}f(r)$, with
\begin{equation}
\label{eq:phipr}
 f(r)=\frac{1}{2}\Big[ 1-\textmd{erf} (2(r-b-\delta)/\delta) \Big],
\end{equation}

where $r$ represents the distance to the microgel center, $\textmd{erf}(x)$ is the error function, $2\delta$ is the thickness of the interface and $\phi_{\text{p}}^\textmd{0}\approx 0.5$ is the accepted value of polymer volume fraction inside a collapsed microgel~\cite{Berndt2005}. This functional dependence leads to the required polymer distribution: It is roughly uniform inside the microgel with $\phi_{\text{p}}(r) \approx \phi_{\text{p}}^0$ for $r < b$, and decays to zero near the external interface, spanning from $r = b$ to $r = b + 2\delta$. For collapsed microgels, the experimental data report a very sharp interface, with a value for the interface half-thickness of about $\delta=1$~nm~\cite{Berndt2005}. We will consider a system temperature of $340$~K, well above the lower critical solution temperature of PNIPAM. This choice ensures that microgel particles composed of this thermo-responsive polymer will be in their collapsed state.

In the initial stage, the microgel is loaded with a certain amount of cargo molecules. Our model does not specify a particular molecule, thus making it broadly applicable to a wide range of non-ionic encapsulated entities, including chemical reactants, biomolecules, and pharmaceutical compounds. It should be noted that charged molecules may involve long-range electrostatic interactions, leading to a very different behavior. As corroborated by a recent simulation study~\cite{Kanduc2021}, the size and shape of the non-ionic molecule contribute to its diffusion coefficient. We proceed under the assumption of a generic cargo molecule, characterized by a hydrodynamic (Stokes) radius $a_w$. This radius is determined from the diffusion coefficient of the molecule in water, $D_w$, using the Stokes--Einstein relation,
\begin{equation}\label{eq:Stokes}
a_w=\frac{k_{\text{B}}T}{6\pi\eta D_w}.
\end{equation}
Here, $T$ is the absolute temperature, $k_{\text{B}}$ is the Boltzmann constant, and $\eta=4.206\times 10^{-4}$ Pa$\cdot$s is the viscosity of water at $T=340$~K.

The two key parameters that strongly control the release kinetics are the effective diffusion coefficient of the cargo molecule inside the microgel, $D_{\text{eff}}$, and the effective interaction between the molecule and the collapsed polymeric network, $u_{\text{eff}}$. On the one hand, $D_{\text{eff}}$ determines the time the molecule needs to diffuse from the initial position inside the microgel to its external interface, which scales as $\tau \sim 1/D_{\text{eff}}$. On the other hand, $u_{\rm eff}$ represents the free energy difference that the molecule must overcome in order to escape to the bulk solution. According to the Arrhenius law~\cite{Laidler1984}, the typical time implied by the molecule to surpass this energy barrier is expected to scale as $\exp (-u_{\text{eff}}/(k_{\text{B}}T))$. Since both quantities describe local properties of the polymer matrix, they depend on the distance to the center of the microgel, $r$. Given the great importance of both functions, we discuss them in detail in the following two sections.

\subsubsection{Effective diffusion coefficient}

The effective diffusion coefficient varies from inside the microgel, denoted by $D^*$, to the corresponding value outside (i.e., in bulk water), given by $D_w$. The switch between these two values at the interface is modeled as
\begin{equation}
D_{\text{eff}}(r)=D_w+(D^*-D_w)f(r),
\end{equation}
where $f(r)$ is given by Eq.~\ref{eq:phipr}. This dependence is illustrated in Fig.~\ref{fig:microgel-diagram}, featuring a sharp increase at the interface.

A distinctive characteristic of collapsed microgels is that water molecules are non-uniformly distributed inside the polymer network, forming irregular nanoscopic clusters, as it usually occurs in similar dense amorphous polymer structures~\cite{Tamai1994,Fukuda1998,Kucukpinar2003}. Experiments~\cite{Ghugare2010,Philipp2014} and computer simulations~\cite{Takeuchi1990,Mueller‐Plathe1991,Sok1992,MuellerPlathe1994} performed on these systems clearly show that, for such dense polymer cross-linked structure, the diffusion of the molecule advances by means of the hopping mechanism. The resulting diffusion coefficient inside such dense polymer networks can be orders of magnitude smaller than in bulk water, of about $D^* \approx (10^{-4}-10^{-2})D_w$~\cite{Zhang2002,kanduc2019free}. In addition, $D^*$ decreases exponentially with the molecule Stokes radius~\cite{Kanduc2018, Kanduc2021}:
\begin{equation}\label{eq:diffCoef}
D^*=D_0e^{-a_w/\lambda},
\end{equation}
where $D_0$ is a reference diffusion constant and $\lambda$ is a characteristic length scale that exclusively depend on the shape of the diffusing molecule.
Schweizer and coworkers~\cite{zhang2017correlated, mei2021activated} have offered a theoretical explanation for the exponential relationship between $D^*$ and the size of the molecule, grounded in the coupled dynamics in dense liquids.

It is important to emphasize that $D^*$ for non-ionic molecules is not affected by the polymer-molecule affinity~\cite{Kanduc2021}. For instance, the value of $D^*$ is the same for polar and non-polar cargo molecules, provided that the size and shape of both molecules are the same in both cases. This is a singular feature of collapsed polymer networks, for which the diffusion of the cargo only depends on its geometrical properties through the steric exclusion effects induced by the polymer chains. The reason for that relies on the fact that, during a jump of the molecule inside such a dense polymer network, the coordination of its solvation shell and the number of contacts with the surrounding polymers are not significantly altered, so the free energy change between the old and the new position is very small compared to obstruction effects caused by the local trapping, which only depends on the size and shape of the molecule~\cite{Kanduc2018, Kanduc2021}.

\subsubsection{Relative shape anisotropy}

The severe restrictions on the motion of the cargo molecules induced by the high polymer concentration inside a collapsed microgel result in privileged directions of diffusion based on the geometry of the cargo molecule. In a previous atomistic study by Kandu\v{c} et al.~\cite{Kanduc2021}, molecules were classified into linear, planar, and spherical groups according to qualitative geometric criteria. This grouping was used to study the behavior of the diffusion coefficient and the potential barrier depth. It was revealed that, for molecules with the same Stokes radii (see Eq.~\ref{eq:Stokes}), linear molecules had the most favorable diffusion, as their shape allowed easy penetration through narrow microgel pores. Planar molecules, displaying intermediate diffusion efficiency, leveraged their shape to diffuse in a disk-like manner. Conversely, spherical molecules faced the most difficulty accessing the microgel pores. Consequently, for a given Stokes radius, $D^*_{\text{linear}} > D^*_{\text{planar}} > D^*_{\text{spherical}}$.

In this work, we adopt a slightly different approach using quantitative criteria to determine diffusion coefficients based on molecular shape. We compute the cargo radius of gyration tensor, $\mathbf{G}$, a method frequently used to characterize the shape of polymers\cite{Solc1971}. Performing the diagonalization of the gyration tensor yields three eigenvalues, $(\alpha_1,\alpha_2,\alpha_3)$.  This intuitively evokes the representation of an ellipsoidal shape, where the eigenvalues correspond to the semiaxes of the ellipsoid. To identify the molecule shape, we calculate a shape descriptor derived from the gyration tensor, called relative shape anisotropy~\cite{Theodorou1985,Arkin2013}:
\begin{equation}
\label{eq:anisotropy}
    \kappa \equiv \frac{3}{2} \frac{\mathrm{Tr}\mathbf{G}^{2}}{(\mathrm{Tr}\mathbf{G})^{2}} = 1 - 3\frac{\alpha_1\alpha_2+\alpha_1\alpha_3+\alpha_2\alpha_3}{(\alpha_1+\alpha_2+\alpha_3)^2},
\end{equation}
which can range from 0 to 1. A linear arrangement of atoms corresponds to $\kappa=1$, a regular planar geometry corresponds to $\kappa=0.25$, and structures of tetrahedral symmetry or higher, such as spheres, correspond to $\kappa=0$\cite{Theodorou1985}. We computed the relative shape anisotropy for each molecule in Table \ref{tab:molecules}, classifying those within a tolerance of $\Delta\kappa=0.1$. 

\begin{figure}[ht!]
\centering
\includegraphics[width=\linewidth]{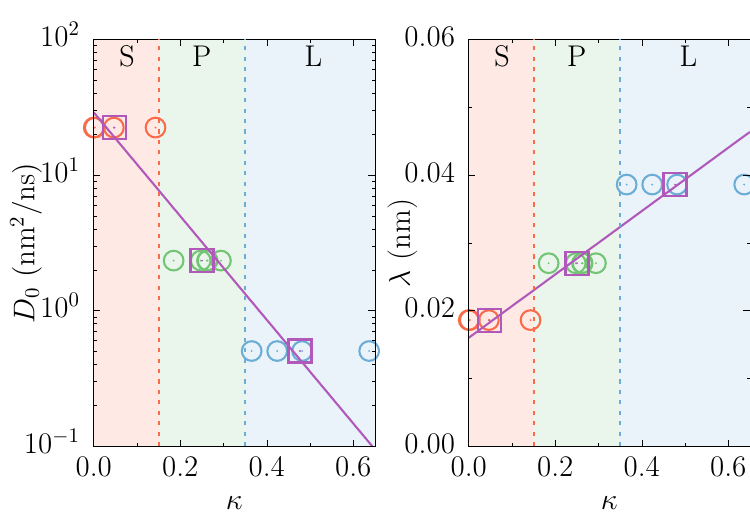}
\caption{Diffusion coefficient prefactor $(D_0)$ and decay length $(\lambda)$ as a function of relative shape anisotropy $(\kappa)$. Each panel is divided into three sections corresponding, from left to right, to the spherical (S), planar (P), and linear (L) shapes of the molecules, respectively.}
\label{fig:kappa-fit}
\end{figure}

Fig.~\ref{fig:kappa-fit} plots the diffusion coefficient prefactor, $D_0$, and decay length, $\lambda$, from Eq.~\ref{eq:diffCoef}, as functions of the relative shape anisotropy, $\kappa$. The diffusion coefficients were taken from a previous simulation work of a collapsed PNIPAM hydrogel~\cite{Kanduc2021}. Molecules are categorized into spherical, planar, or linear groups based on their $\kappa$ values, and within each group, average values $\langle D_0\rangle$ and $\langle\lambda\rangle$ are calculated. The data are then fitted using the following functions:
\begin{align}
\lambda = & \ m_{1}\kappa + n_{1}\label{eq:lambda-kappa-fit},\\
D_{0} =   & \ n_{2}\ \mathrm{e}^{m_{2}\kappa}\label{eq:D0-kappa-fit}.
\end{align}
The left panel of Fig.~\ref{fig:kappa-fit}, employing a linear-log scale, reveals an exponential relationship between $D_0$ and $\kappa$, outlined in Eq.~\ref{eq:D0-kappa-fit}. The right panel displays a linear correlation between $\lambda$ and $\kappa$, as described by Eq.~\ref{eq:lambda-kappa-fit}. The values of the fitting parameters for each plot are compiled in Table \ref{tab:fit-parameters}. These parameters allow us to compute $D_0$ and $\lambda$, and therefore the value of $D^{*}$ for every non-ionic molecule by using Eq.~\ref{eq:diffCoef}, just by computing the corresponding shape anisotropy, $\kappa$.
\begin{table}[ht!]
\caption{Numerical values for the fitting parameters obtained from the fits in Fig. \ref{fig:kappa-fit}. Parameters $(m_{1},n_{1})$ are derived from the linear fit of $\lambda$ as a function of $\kappa$. Parameters $(m_{2},n_{2})$ result from the linear-log fit of $D_{0}$ as a function of $\kappa$.}
\label{tab:fit-parameters}
\begin{tabular}{lll}
\hline
\begin{tabular}[c]{@{}l@{}}Fitting\\ parameter\end{tabular} & Value   & \begin{tabular}[c]{@{}l@{}}Standard\\ error\end{tabular} \\
\hline
$m_{1}$ (nm)            & \ \ \ \ 0.047  & \ \ \ 0.003  \\
$n_{1}$ (nm)            & \ \ \ \ 0.0160 & \ \ \ 0.0009 \\
$m_{2}$                 & \ \ \ -8.8      & \ \ \ 1.3     \\
$n_{2}$ (nm$^2$/ns)     & \ \ \ \ 3.37    & \ \ \ 0.39    \\     
\hline
\end{tabular}
\end{table}
\subsubsection{Effective molecule-microgel interaction}

Transferring the molecule from the bulk solution (water phase) into the microgel (polymer phase) is characterized by the free energy difference, $\Delta G$. In a similar fashion as we modeled $\phi_{\text{p}}(r)$ and $D^*(r)$, we assume the radial variation of the effective interaction as  $u_{\text{eff}}(r)=\Delta G f(r)$. In contrast to the diffusion coefficient, $\Delta G$ depends not only on the size and shape of the molecule but also on its chemical nature, particularly its polarity. Thus, the molecule's affinity to the polymer matrix is strongly influenced by its hydrophobic or hydrophilic nature. In a simple phenomenological approach, this energetic contribution scales very well with the molecule's surface area, and can be written as~\cite{kanduc2019free, Kanduc2021}
\begin{equation}\label{eq:effPot}
\Delta G = \Delta G_0 + \gamma_0 4\pi a_\text{AS}^2.
\end{equation}
In the second term, $4\pi a_\text{AS}^2$ is the solvent-accessible surface area of the molecule, expressed with an equivalent spherical radius, $a_\text{AS}$. Note that $a_\text{AS}$ is strongly linked to the Stokes radius, and for a wide range of diverse molecules, it can be expressed as $a_\text{AS}=a_w+0.233$~nm~\cite{Kanduc2021}. The coefficient $\gamma_0$ can be interpreted as the disparity between the surface tensions of the molecule with the polymer and the molecule with water. The first term in Eq.~\ref{eq:effPot}, $\Delta G_0$, reflects the chemical nature of the molecule, correlating with the number of polar groups present within it.
\begin{table}[ht!]
\caption{Interaction free energies and diffusion coefficients, taken 
from a previous simulation work~\cite{Kanduc2021}, which are investigated in this study. Molecules have been classified based on the relative shape anisotropy, as defined by Eq.~\ref{eq:anisotropy}.}
\resizebox{\linewidth}{!}{\begin{tabular}{l|l|r|c}
\hline
Molecule & Symbol & $\beta \Delta G$ & $D^*/D_w\times10^{4}$ \\
\hline
\multicolumn{4}{c}{Linear} \\
\hline
Nitrophenol & NP & $-7.90$ & $1.5$ \\
Pentane & Pe & $-6.22$ & $4$ \\
Pentanol & PeOH & $-4.44$ & $3.2$ \\
\hline
\multicolumn{4}{c}{Planar} \\
\hline
Nitrobenzene & NB & $-7.54$ & $3.6$ \\
Phenol & Ph & $-6.19$ & $2.4$ \\
Benzene & B & $-4.88$ & $3.7$ \\
Propanol & PrOH & $-2.07$ & $6.3$ \\
Methanol & MeOH & $-0.36$ & $27$ \\
\hline
\multicolumn{4}{c}{Spherical} \\
\hline
Tetraclhoromethane & CCl$_4$ & $-8.75$ & $1.12$ \\
Neopentane & NPe & $-8.43$ & $0.77$ \\
Ethane & Et & $-3.44$ & $17$ \\
Methane & Me & $-1.88$ & $71$ \\
Argon & Ar & $-1.51$ & $135$ \\
Neon & Ne & $-0.20$ & $810$ \\
Helium & He & $0.19$ & $1970$ \\
\hline
\end{tabular}}\label{tab:molecules}
\end{table}

\subsection{Dynamical Density Functional Theory}

To investigate the non-equilibrium diffusive release of cargo molecules, we make use of classical Dynamical Density Functional Theory (DDFT)~\cite{Marconi1999,Archer2004,Wu2007,Vrugt_2023}. DDFT is a theoretical framework for the time evolution of the one-body density of a fluid. It extends the original framework of the density functional theory (DFT), which was designed for equilibrium systems~\cite{Hansen2013,Evans1992,Ebner1976,Ebner1977}, to address non-equilibrium cases. This method can be used to describe how an initial non-equilibrium density profile of molecules evolves in time in the presence of an external potential exerted by the microgel. In particular, it provides the local concentration of the cargo on position $\mathbf{r}$ at time $t$ during the release process, $\rho_{\text{c}}(\mathbf{r},t)$, starting from an initial distribution, $\rho_{\text{c}}(\mathbf{r},0)$.

In contrast to the usually employed diffusion equation, DDFT has three important improvements. First, the diffusion of the cargo takes into account the interaction of the molecule with the external field exerted by the polymer network of the microgel, $u_{\text{eff}}(r)$. Second, the method also considers the position dependence of the diffusion coefficient, $D_{\text{eff}}(r)$. Indeed, this is precisely the case when the molecule travels from the interior of the microgel to the bulk solution, where the diffusion constant changes from $D^*$ to $D_w$. Finally, DDFT also incorporates the effect of the molecule--molecule interactions by means of the corresponding excess free energy.

Within the DDFT approach, the cargo concentration obeys the following continuity equation~\cite{Marconi1999,Wu2007}: 
\begin{equation}
\label{eq:DDFT1}
\frac{\partial \rho_{\text{c}}(\mathbf{r},t)}{\partial t}=-\nabla \cdot J_{\text{c}}(\mathbf{r},t),
\end{equation}
where $J_{\text{c}}(\mathbf{r},t)$ denotes the space and time-dependent diffusive flux, given by 
\begin{equation}
\label{eq:DDFT2}
J_{\text{c}}=-D_{\text{c}}(\mathbf{r})\Big[ \nabla \rho_{\text{c}}+\rho_{\text{c}}\beta\nabla \big( u_{\text{eff}}(\mathbf{r}) + \mu_{\text{c}}^\textmd{ex}(\mathbf{r},t) \big) \Big],
\end{equation}
where $\beta = 1/(k_{\text{B}}T)$. The excess non-equilibrium chemical potential, $\mu_{\text{c}}^\textmd{ex}(\mathbf{r},t)$, gathers the effect of molecule--molecule interactions. In general, it is obtained from the functional derivative of the equilibrium excess free energy of the molecules~\cite{Evans1992,Hansen2013}. Here, we use a simple mean-field prescription to account for excluded-volume interactions, given by the Carnahan--Starling expression~\cite{Hansen2013}:
\begin{equation}\label{muc}
\beta \mu_{\text{c}}^\textmd{ex}(\mathbf{r},t) =\phi_{\text{c}}(\mathbf{r},t)\frac{8-9\phi_{\text{c}}(\mathbf{r},t)+3\phi_{\text{c}}(\mathbf{r},t)^2}{
	[1-\phi_{\text{c}}(\mathbf{r},t)]^3},
\end{equation}
where $\phi_{\text{c}}(r,t)=\frac{4}{3}\pi a_w^3\rho_{\text{c}}(r,t)$ is the local volume fraction of molecules.

Eqs.~\ref{eq:DDFT1} and \ref{eq:DDFT2} must be solved numerically for spherical symmetry. Three boundary conditions are required. The first one establishes the initial distribution of cargo molecules. In our case,  they are initially distributed uniformly inside the microgel particle,
\begin{equation}
\label{eq:t0}
\rho_{\text{c}}(r,t=0)=\begin{cases}
\rho_{\text{c}}^0 & r \le b \\
	0 & r>b
\end{cases}.
\end{equation}
Second, the diffusive flux in the center of the microgel must be zero due to the spherical symmetry of the system, 
\begin{equation}
J_{\text{c}}(r=0,t)=0 \quad \forall t.
\end{equation}
Finally, for a very diluted suspension of microgel particles, we can assume that the volume of the bulk solution surrounding the microgel is very large, such that the cargo concentration far away from the microgel is zero. This implies the following absorbing boundary condition:
\begin{equation}
\rho_{\text{c}}(r \to \infty,t)=0 \quad \forall t.
\end{equation}
For practical reasons, we consider that a distance $R=20b$ is large enough to apply this boundary condition, such that $\rho_{\text{c}}(r=R,t)=0$.

In our calculations, cargo molecules are assumed to be released to the bulk solution when they reach the outer interface of the microgel, that is, when $r>b+2\delta$. With this prescription, the fraction of released cargo is given by
\begin{equation}\label{eq:f_rel}
f_{\text{rel}}(t)=1-\frac{4\pi}{N_0}\int_{0}^{b+2\delta}r^2\rho_{\text{c}}(r,t)\mathrm{d}r,    
\end{equation}
where $N_0$ is the number of molecules encapsulated inside the microgel in the initial state, $N_0=\frac{4}{3}\pi b^3\rho_{\text{c}}^0$. We define the half-release time, $\tau_{1/2}$, as the time required to release 50\% of the encapsulated cargo, i.e.
\begin{equation}
\label{tau12def}
f_{\text{rel}}(\tau_{1/2})=0.5.
\end{equation}
As we will see later, $\tau_{1/2}$ holds immense significance as it consolidates the most pertinent details regarding the release kinetics into a singular parameter.  In fact, the knowledge of the scaling $\tau_{1/2}$ with $D^*$ and $\Delta G$ is especially of vital help to estimate extremely slow release kinetics where the integration of the DDFT equation can involve prohibitively long calculations. 

In addition to $\tau_{1/2}$, we can define the mean release time of the encapsulated molecules, given by
\begin{equation}
\label{meanreltime}
\bar{\tau}=\int_0^{\infty}t\Big( \frac{\mathrm{d}f_{\text{rel}}}{\mathrm{d}t}\Big) \mathrm{d}t.
\end{equation}

To solve the time-dependent DDFT differential equation, distances were scaled by $l_0=1$~nm, and time by $\tau_0=l_0^2/D_w$. In order to shorten the computation time of the numerical resolution, a non-uniform spatial grid was used to sample the distance $r$. We used a very narrow grid size of $\Delta r_{\text{min}}=0.02l_0$ at the microgel interface, where the gradients of $D_{\text{eff}}(r)$ and $u_{\text{eff}}(r)$ are larger, whereas a thicker size intervals $\Delta r_{\text{min}}=0.5l_0$ were employed in the regions inside and outside the microgels. On the other hand, a time step of $\Delta t=10^{-4}\tau_0$ was used in all the calculations. This value is smaller than $(\Delta r_{\text{min}})^2/(2D_0)$, thus preventing the occurrence of non-physical sawtooth instabilities.

\section{Results and discussion}\label{sec:results}
In this section we make use of the DDFT theoretical framework described above to study the release kinetics of the molecules encapsulated within the collapsed microgel. This technique provides the time evolution of the cargo's density profiles, enabling us to determine the fraction of released molecules and the characteristic release time. In all the cases studied here, the initial encapsulated cargo concentration contained inside the microgel is $\rho_{\text{c}}^0=0.01$~M.

\subsection{Spatio-temporal evolution of cargo molecules}
\begin{figure*}[ht!]
\centering
\includegraphics[width=\linewidth]{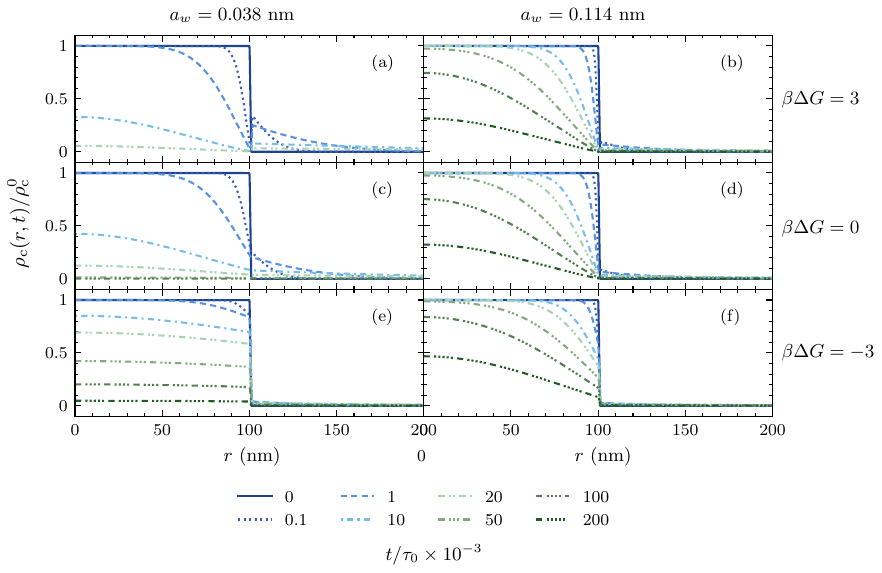}
\caption{Spatial evolution of the density of cargo molecules normalized by their initial density inside the microgel; different lines represent different snapshots in time. Each column depicts the results for a given diffusion coefficient (or molecular size). In panels (a), (c), and (e), the Stokes radius corresponds to that of helium, that is $a_w = 0.038$~nm. Additionally, in panels (b), (d), and (f) the cargo molecular size is restricted to the methane Stokes radius, which is $a_w = 0.114\ \rm nm$. Every row of panels represents a specific value of the potential barrier $\beta\Delta G$: (a) and (b) to $\beta \Delta G = 3$, (c) and (d) to $\beta \Delta G = 0$, and (e) and (f) to $\beta \Delta G = -3$. In all cases, the geometry of the molecule is spherical, $\rho_{\rm c}^0 = 0.01$~M, and $\delta = 1$~nm.}
\label{fig:density-evo}
\end{figure*}
Fig.~\ref{fig:density-evo} shows the time evolution of the local density of cargo molecules normalized by the initial density inside the microgel, $\rho_{\text{c}}(r,t)/\rho_{\text{c}}^0$, for two molecular sizes and three molecule--microgel interaction free energies, covering attractive, neutral, and repulsive polymer networks ($\beta \Delta G=-3$, $0$ and $3$, respectively). We consider the particular case of spherical molecules ($\kappa=0$), although similar results are found for other shapes. The results are organized so that each column represents a specific molecule size. From these graphs, it becomes clear that the larger molecules are released at a slower rate than their smaller counterparts. This inverse relation between molecular size and release rate can be attributed to the exponential decay of the diffusion coefficient with the molecule size, $a_w$.

Fig.~\ref{fig:density-evo} also shows the influence of various free energy values on the release process. Each row of panels corresponds to a specific value of $\beta \Delta G$. We observe that attractive microgels remarkably slow down the release process, which can be observed in the difference between concentration profiles over time when comparing $\beta \Delta G = -3$ to other potential barriers. Important differences are observed between the time-dependent density profiles in repulsive and attractive polymer networks. For repulsive networks such as $\beta \Delta G=+3$ the cargo molecules are energetically expelled from the microgel, leading to a dynamic depletion of molecules close to the interface of the microgel (see Fig.~\ref{fig:density-evo}(a) and (b)). Conversely, for attractive networks ($\beta \Delta G=-3$), molecules are retained inside the microgel because they need to overcome a free energy step-edge barrier to escape from the interior of the microgel to the bulk solution, giving rise to a more uniform distribution of molecules, especially for small cargo molecules, as seen in Fig.~\ref{fig:density-evo}(e).

\subsection{Time evolution of the fraction of released molecules}
\begin{figure}[ht!]
\centering
\includegraphics[width=\linewidth]{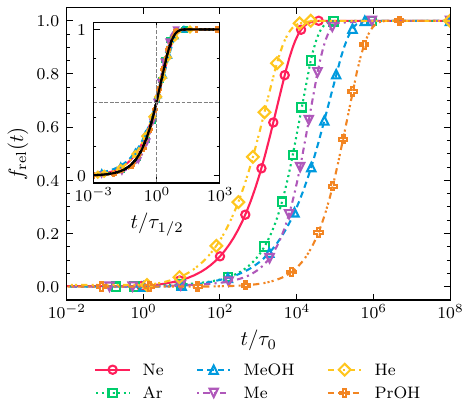}
\caption{Time evolution of the fraction of released cargo for different molecules. The inset represents the same dataset, with the time axis normalized by the half-release time. A reference grid helps emphasizing how all the renormalized curves align at $(1,0.5)$. In all cases, the used parameters are $\rho_{\rm c}^0 = 0.01$~M,  $\delta = 1$~nm, and $b=50$~nm. The black solid line shows the fit provided by Eq.~\ref{CDF}.}
\label{fig:cargo-evo}
\end{figure}
Integrating the density profiles $\rho_{\text{c}}(r,t)$ in Eq.~\ref{eq:f_rel} leads to the time-dependent fraction of release cargo, $f_{\text{rel}}(t)$. In Fig.~\ref{fig:cargo-evo} we depict $f_{\text{rel}}(t)$ for a specific set of molecules, covering small and large sizes and different molecule--microgel interaction free energies. As observed, $f_{\text{rel}}(t)$ exhibits a profile consistent with a cumulative distribution function. A very important quantity that characterizes the release process is the so-called half-release time, $\tau_{1/2}$, defined as the time at which $50$\% of the molecules have been released. Notably, while all the curves exhibit analogous profiles, the primary variance among them is attributed to the $\tau_{1/2}$ value, arising from the particular release dynamics of each molecule. Indeed, the curve corresponding to the diffusive release of large-sized molecules that are strongly attracted to the microgel is shifted to longer times compared to small molecules weakly attracted to the microgel, because the former ones diffuse slower and also need to surpass a higher free energy barrier to reach the bulk solution.

Normalizing the time by the respective half-release times for each molecule allows all these curves to converge into a master curve, as shown in the inset of Fig.~\ref{fig:cargo-evo}. The overlap of all of the release profiles upon this normalization suggests a potential universal feature of the release process, which is scalable across different molecular types. In addition, this scaling demonstrates that $\tau_{1/2}$ is enough to fully predict the release kinetics of any molecule. The percentage of released cargo molecules, of various shapes and sizes can be described by a common Weibull cumulative distribution function~\cite{Casault2008,Wang2017}
\begin{equation}\label{CDF}
    f_\mathrm{rel}(t) = 1-\exp{\left[ -\left(\chi t / \tau_{1/2}\right)^\nu\right]},
\end{equation}
where $\chi$ and $\nu$ are the fitting parameters controlling the shape of the function, with values $\chi = 0.64\pm0.04$ and $\nu = 0.79\pm 0.13$, as determined from fitting the DDFT results. From this cumulative distribution, the mean release time of the molecules can be calculated using Eq.~\ref{meanreltime}, resulting in
\begin{equation}
{\bar\tau}=1.787\tau_{1/2}.
\end{equation}

\subsection{Effect of the diffusion coefficient and molecule--microgel interaction free energy}
\begin{figure*}[ht!]
\centering
\includegraphics[width=\textwidth]{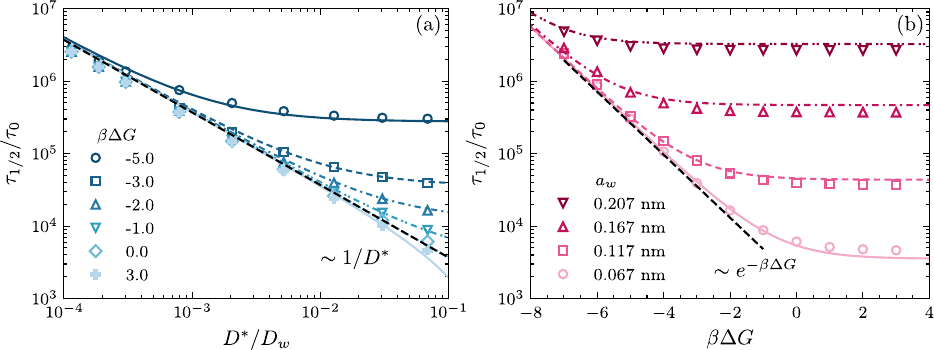}
\caption{Normalized half-release time as a function of the diffusion coefficient and the molecule--microgel interaction free energy. Symbols represent results obtained from DDFT, while lines correspond to theoretical predictions from Eqs.~\ref{taubarfinal} and \ref{eq:tau-half-bar}. Fig.~\ref{fig:diffCoef-DeltaG}(a) displays $\tau_{1/2}/\tau_{0}$ plotted against the diffusion coefficient inside the microgel $D^*$ normalized by the value in bulk, $D_w$, for various values of the microgel-molecule interaction free energy, $\beta \Delta G=\{-5, -3, -2, -1, 0, 3\}$. Conversely, Fig.~\ref{fig:diffCoef-DeltaG}(b) illustrates $\tau_{1/2}/\tau_{0}$ as a function of $\beta\Delta G$ for different molecular sizes, given by $a_w=\{0.067, 0.117, 0.167, 0.208\}$~nm. For all presented cases, the molecule has a spherical geometry, with $\rho_{\rm c}^0 = 0.01$~M, $b=100$~nm, and $\delta = 1$~nm.}
\label{fig:diffCoef-DeltaG}
\end{figure*}
Recognizing $\tau_{1/2}$ as the inherent time scale of the release, it becomes imperative to thoroughly explore its dependency on $D^*$ and $\Delta G$ as the main material parameters of our system. The goal is to find an analytical expression $\tau_{1/2}(D^*,\Delta G)$ capable of universally characterizing the release kinetics of any non-ionic molecule from collapsed microgels.

Fig.~\ref{fig:diffCoef-DeltaG} illustrates how the half-release time, derived from solving the DDFT equation, correlates with the effective diffusion coefficient $D^*$ for different $\Delta G$ values, including both repulsive and attractive interactions.
For molecules with small diffusion coefficient the release of cargo is primarily governed by diffusion. This observation aligns with well-established diffusion principles of Brownian particles, which state that the mean time to travel a fixed distance is inversely proportional to the diffusion coefficient, i.e. $\tau \sim 1/D^*$. We emphasize that, in this diffusion-limited regime, the effect of $\Delta G$ is negligible because the release kinetics are completely controlled by the time the molecules need to diffuse to the external interface of the microgel. The time to surpass the microgel interface is only a minor correction in this case. Consequently, all curves with different $\Delta G$ collapse onto a common form in this regime. The use of a normalized release time $\tau_{1/2}/\tau_0$ enhances the visibility of this collapse since it removes the dependence of the half-release time with $D_w$.

As $D^*$ increases, indicative of smaller particle sizes, we observe a departure from the law $\tau_{1/2} \sim 1/D^*$ when examining attractive microgels. Indeed, beyond a certain point, the effects of negative $\Delta G$ over small cargo become significant in determining their release time. The reason for this relies on the fact that, for such fast molecules, the diffusive time inside the microgel becomes smaller. Therefore, the role played by the interaction free energy starts to be relevant, as the molecule needs to spend a significant time to overcome the attraction free energy barrier located at the microgel external surface for the case of attractive polymer networks. In other words, the molecule release becomes a reaction-limited process.

This is closely related to the dependence of the release time with the effective potential, shown as symbols in Fig.~\ref{fig:diffCoef-DeltaG}(b). When considering negative values of $\Delta G$, we find that the release time increases as $\tau_{1/2} \sim \exp{(-\beta \Delta G)}=\exp{(\beta | \Delta G |)}$ (see black dashed line), suggesting that the release kinetics follows an Arrhenius law. However, when the free energy is positive, its effects become negligible, indicating a saturation or threshold effect, beyond which the energetic profile exerted by the microgel does not affect the release time, leading to the diffusion-limited regime, in which  $\tau_{1/2}$ is independent of $\Delta G$. In this realm of interactions, the $\tau_{1/2} \sim 1/D^*$ tendency is manifested through the logarithmic shifts for different molecular sizes (see Eq.~\ref{eq:diffCoef}).

Clearly, the existence of these two kinetic regimes indicates the presence of two major processes involved in the release kinetics: the diffusion through the microgel and the surpass of the free energy barrier. Later, we will deepen into the role of these two contributions and propose an analytical model for $\tau_{1/2}$ in terms of both physical parameters.

It is important to remark that, as indicated by Eqs.~\ref{eq:diffCoef} and \ref{eq:effPot}, the value of $D^*$ and $\Delta G$ are not independent for a particular molecule. In other words, it is not possible to fix $a_w$ (i.e., $D^*$) and then arbitrarily vary $\Delta G$ and vice versa. In this sense, the plots in Fig.~\ref{fig:diffCoef-DeltaG} do not precisely represent results for a particular cargo molecule. Instead, they provide general physical insights useful for understanding the overall dependence of $\tau_{1/2}$ with our model parameters.

\subsection{Effect of the microgel size}
\begin{figure*}[ht!]
\centering
\includegraphics[width=\textwidth]{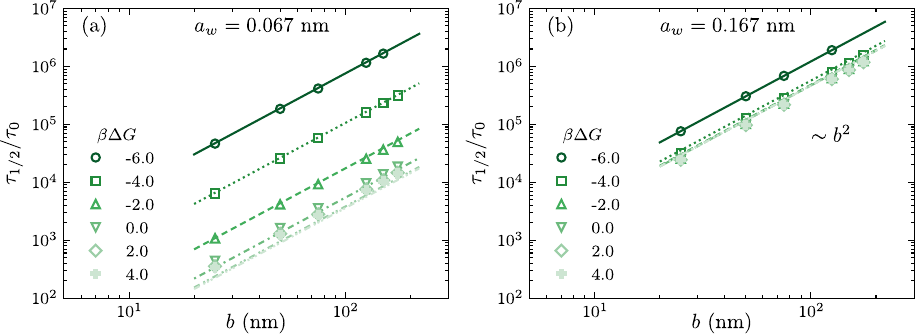}
\caption{Half-release time, $\tau_{1/2}$, against the radius of the microgel $b$. The symbols represent data obtained by numerical integration of the DDFT equations, whereas the lines correspond to the theoretical predictions given by Eqs.~\ref{taubarfinal} and \ref{eq:tau-half-bar}. Left panel (Fig.~\ref{fig:tau-size}(a)) displays the case in which $a_w = 0.067\ \rm nm$, a value that is identified with Neon. Right panel (Fig.~\ref{fig:tau-size}(b)) plots the same results, but for $a_w = 0.167$~nm. Each plot shows results for several values of $\beta \Delta G$, which extend from $-6$ to $4$. In all cases, the geometry of the molecule is spherical, $\rho_{\rm c}^0 = 0.01 \, \rm M$, and $\delta = 1 \, \rm nm$.}
\label{fig:tau-size}
\end{figure*}
Fig.~\ref{fig:tau-size} dives into the relationship between the half-release time $\tau_{1/2}$ and the radius of the collapsed microgel, $b$. The data obtained for different values of $a_w$ and $\Delta G$ (shown as symbols) reveals that there is a direct proportionality between the release time and the square of the microgel's radius, $\tau_{1/2} \sim b^2$. Clearly, a larger microgel requires a longer release time, since the encapsulated molecules have to diffuse a larger distance to reach the interface and escape to the bulk solution.  This correlation is, in fact, related to a fundamental equation in diffusion studies known as the mean square displacement equation, $\langle \Delta\mathbf{r}^2\rangle = 6Dt$.

We can also appreciate the effect that different molecular sizes and free energy potential depths have on the release time. Repulsive and non-interacting microgels ($\Delta G \ge 0$) behave in somewhat the same way: the transfer free energy has a very weak effect on the release time. However, for attractive microgels ($\Delta G < 0$) the release time scales exponentially with the attraction strength, as observed in the log-scaled constant shifts of the release time in Fig.~\ref{fig:tau-size}(a). In addition, Fig.~\ref{fig:tau-size}(b) shows that the influence that cargo size has over the kinetics is to universalize the release time all across the potential range once the molecular radii become sufficiently large. That is, diffusion on the microgel becomes a matter of size and not interaction. This is something that has been already noted in Fig.~\ref{fig:diffCoef-DeltaG}(a), where larger cargo shared a common $\tau_{1/2}$ regardless of the value of $\Delta G$.

\subsection{Analytical expression for the half-release time}

In order to gather all the scaling trends into a unique analytical expression and provide a physical interpretation of the DDFT results, we calculate the mean-first passage time (MFPT), defined as the mean time that the molecules contained inside the microgel reach the bulk solution for the first time~\cite{Klein1952,Lifson1962,Weiss1967,Szabo1980}. Since the concentration of cargo molecules considered in our DDFT calculations is low everywhere, as a first approximation the molecule--molecule interactions can be neglected, so the release process can be modeled as a diffusion problem of molecules through the effective potential induced by the microgel, $u_{\text{eff}}(r)$. In this limit, the DDFT equations simplify to the well-known Smoluchowski equation. We consider a suspension of microgels, in which each microgel occupies, on average, a volume $V$. This volume $V$ is the total system volume divided by the number of microgel particles. We approximate, as in our DDFT calculations, this volume $V$ as a sphere of radius $R$, such that the volume fraction of microgels in the suspension is $\varphi = (b/R)^3$. We apply  an absorbing boundary condition at $r=R$, which means that the molecules that reach this distance disappear from the system. We now turn our attention to a molecule initially located at a distance $r=s$ from the center of the microgel. We are interested in determining the average time spent by this molecule to exit the designated volume, essentially reaching the surface of the spherical cell. The derivation, shown in detail in the \hyperref[sec:appendix]{Appendix}, leads to the following expression for the MFPT~\cite{Lifson1962,Szabo1980}:
\begin{equation}
\bar{\tau}(s)=\int_{s}^R \mathrm{d}r\frac{e^{\beta u_{\text{eff}}(r)}}{r^2D_{\text{eff}}(r)}\int_0^r \mathrm{d}r^\prime {r^\prime}^2e^{\beta u_{\text{eff}}(r^\prime)}.
\end{equation}
In order to perform the integrals and obtain an analytical expression for $\bar{\tau}(s)$, we approximate the interface of the microgel as a sharp boundary, such that $u_{\text{eff}}(r)=\Delta G$ for $r \le b$, and $u_{\text{eff}}(r)= 0$ for $r > b$. Analogously, the effective diffusion coefficient is simplified to $D_{\text{eff}}(r) = D^*$ for $r \le b$ and $D_{\text{eff}}(r) = D_w$ for $r > b$. Both assumptions are fully justified since collapsed microgels have a very narrow interface. With all these simplifications, we find the MFPT of a molecule located at $r = s$ as
\begin{equation}
\bar{\tau}(s)=\frac{b^2-s^2}{6D^*}+\frac{b^3}{3D_w}\Big( \frac{1}{b}-\frac{1}{R}\Big)(e^{-\beta \Delta G}-1) + \frac{R^2-b^2}{6D_w}.
\end{equation}
To calculate the average MFPT of all the encapsulated molecules, we have to perform a second averaging, accounting for the  uniform distribution of molecules within the microgel in the initial stage, so $\bar{\tau}=\int_0^b s^2 \bar{\tau}(s)\mathrm{d}s/\int_0^b s^2 \mathrm{d}s$, leading to
\begin{equation}
\label{taubar}
\bar{\tau}=\frac{b^2}{15D^*}+\frac{b^3}{3D_w}\Big( \frac{1}{b}-\frac{1}{R}\Big)(e^{-\beta \Delta G}-1) + \frac{R^2-b^2}{6D_w}.
\end{equation}
Each term on the right-hand side of Eq.~\ref{taubar} has a clear interpretation. The first term is the
average time of diffusion of the molecules inside the microgel to reach the microgel interface at $r = b$. The second term represents the time to overcome the free energy well $\Delta G$. Finally, the third term is the time spent by the molecules to diffuse from $r = b$ to reach the bulk at $r = R$.

For the comparison between this theoretical prediction and the DDFT calculations we only need to retain the first two terms of Eq.~\ref{taubar}, as they provide the time spent by the molecules to escape outside the microgel. Then, taking a very diluted suspension of microgels ($R\gg b$) we obtain the MFPT for the molecules to escape from the microgel:
\begin{equation}
\label{taubarfinal}
\bar{\tau}=\frac{b^2}{15D^*}+\frac{b^2}{3D_w}(e^{-\beta \Delta G}-1).
\end{equation}
Finally, the corresponding half release time will be given by
\begin{equation}
    \tau_{1/2}=\bar{\tau}/1.787.
    \label{eq:tau-half-bar}
\end{equation}
Eq.~\ref{taubarfinal} gathers all the expected scalings of the release time with $b$, $D^*$, and $\Delta G$. Our equation aims to capture the intricate dynamics involved in the release kinetics of molecular cargo from collapsed microgels.
This formula includes two terms that highlight the two fundamental forces at play. The first term mirrors the influence of pure diffusion on the release kinetics. This term embodies the classic principles of diffusive transport, suggesting that the larger the microgel, the longer the release time, while faster diffusion coefficients lead to shorter release times. The second term introduces the influence of the free energy profile of the microgel-cargo system.

As observed in Fig.~\ref{fig:diffCoef-DeltaG} and \ref{fig:tau-size}, the analytical estimate given by Eq.~\ref{taubarfinal}, represented by lines, demonstrates remarkable predictive accuracy across diverse conditions in our study. Particularly, this finding enables us to predict the release of cargo $f_{\rm rel}(t)$ portrayed in Fig.~\ref{fig:cargo-evo} by fusing Eq.~\ref{CDF} with Eq.~\ref{taubarfinal}. Consequently, the fraction of released cargo can be accurately solved at any given time upon knowing the diffusion coefficient, the interaction microgel-molecule interaction free energy, and the radius of the microgel. We believe that this equation presents a powerful tool for researchers. It not only offers a conceptual framework for contemplating release kinetics but also provides a way for making preliminary predictions that can be refined using experimental data. We also believe that the equation can serve as a catalyst for further advancements in the field, fostering a more profound understanding of microgel behavior and potentially contributing to progress in applications such as drug delivery systems.

\section{Conclusions}\label{sec:conclusions}
In this study, we have comprehensively characterized the non-equilibrium diffusive release kinetics of molecules with varying shapes and sizes from collapsed microgels. Our approach uses Dynamical Density Functional Theory, a theoretical framework specifically designed to incorporate diffusion coefficients and transfer free energies of subnanometer-sized molecules in collapsed PNIPAM polymers in water, obtained from atomistic molecular dynamics simulations.

We have uncovered a universal behavior in the release dynamics of different molecules, which can be scaled using a single parameter: the half-release time, $\tau_{1/2}$. In addition, our calculations show that the value of $\tau_{1/2}$ is governed by only two basic material parameters, namely the diffusion coefficient within the polymer network, $D^*$, and the free energy of interaction between the molecule and the microgel,  $\Delta G$. Notably, both of these variables are shaped by the geometry  of the transported molecule. Linear-shaped molecules navigate through the microgel more efficiently compared to their spherical counterparts. Geometry even appears to dictate the preferred directions of molecular movement.

We have consolidated these key variables into a single analytical formula for $\tau_{1/2}$, which has shown excellent agreement with our DDFT calculations. We have observed a clear distinction between collapsed microgels with attractive and repulsive characteristics. The former type significantly extends the release time, governed by the free energy barrier $\tau_{1/2}\sim \exp(-\Delta G/k_{\text{B}}T)$—this effect is particularly pronounced with smaller cargo molecules, wich diffuse fast through the polymer matrix. In contrast, larger molecules align more closely to the diffusion theory for their release, thus $\tau_{1/2}\sim 1/D^*$. In either case, the relationship between the microgel particle size and $\tau_{1/2}$ strictly follows a well-known power law of $\tau_{1/2}\sim b^2$, with $b$ representing the microgel radius.

In subsequent studies, we plan to study the release kinetics from swollen microgels, anticipating that diffusion within these structures will differ from that in collapsed microgels. The ultimate goal is to establish a theoretical framework capable of predicting cargo diffusion within microgel particles for all swelling states.

\section*{Appendix. Calculation of the mean first passage time}\label{sec:appendix}
In the limit of negligible molecule--molecule interactions, the DDFT formalism converges to the Smoluchowski equation for the time-dependent probability density, $\rho(\mathbf{r},t)$, defined as the probability density of finding a Brownian particle (molecule) at position $\mathbf{r}$ at time $t$~\cite{Lifson1962,Szabo1980}
\begin{equation}
\frac{\partial \rho}{\partial t}=\nabla \cdot \big[ D(\mathbf{r})\nabla \rho + D(\mathbf{r})\rho \beta \nabla u(\mathbf{r}) \big],
\end{equation}
where $D({\mathbf{r}})$ is the position-dependent diffusion coefficient and $u(\mathbf{r})$ is the external interaction potential acting on the Brownian molecules. If at time $t=0$ the molecules are localized in a very small region at position $\mathbf{r}$, the probability density can be denoted as $\rho(\mathbf{r}^\prime, t | \mathbf{r}, 0)$, which satisfies the initial condition 
$\rho(\mathbf{r}^\prime, 0 | \mathbf{r}, 0)=\delta (\mathbf{r}-\mathbf{r}^\prime)$.

We assume that the molecules are contained within a volume $V$. The molecules can escape from this volume by reaching its surface $\Sigma$. This means that the system has an absorbing surface, with $\rho(\mathbf{r} \in \Sigma,t)$. The probability that a given molecule initially located at position $\mathbf{r}$ inside $V$ at $t = 0$ escapes outside this volume at time $t$ will be denoted by $W(r,t)$, given by
\begin{equation}
W(\mathbf{r},t)= 1-\int_V \rho(\mathbf{r}^\prime, t |  \mathbf{r}, 0)\mathrm{d}\mathbf{r}^\prime .   
\end{equation}

This new function $W$ satisfies several conditions. On the one hand, if the point $\mathbf{r}$ is located at the surface of $V$, then $W(\mathbf{r},t) = 1$. On the other hand, if $\mathbf{r}$ is a point inside of $V$, then $W(\mathbf{r},0) = 0$ and $W(\mathbf{r},\infty) = 1$. $W(\mathbf{r},t)$ obeys the following partial differential equation~\cite{Lifson1962,Szabo1980}:
\begin{equation}
\label{W}
\frac{\partial W}{\partial t}=\nabla \cdot \big[ D(\mathbf{r})\nabla W \big] - D(\mathbf{r}) \beta \nabla u(\mathbf{r}) \cdot \nabla W.
\end{equation}
The probability that a molecule located at $\mathbf{r}$ at time $t = 0$ arrives to the surface of the volume and escapes outside between $t$ and $t+\mathrm{d}t$ is
\begin{equation}
W(\mathbf{r},t+\mathrm{d}t)-W(\mathbf{r},t)=\frac{\partial W(\mathbf{r},t)}{\partial t}\mathrm{d}t.    
\end{equation}
Therefore, the average time for a molecule located at $\mathbf{r}$ at time $t = 0$ to arrive at the surface is given by the following integral,
\begin{equation}
\bar{\tau}(\mathbf{r})=\int_{0}^{\infty} t \frac{\partial W(\mathbf{r},t)}{\partial t} \mathrm{d}t.
\end{equation}
This is precisely the so-called mean first passage time (MFPT). We can deduce the differential equation that satisfies the MFPT by applying a second-time derivative in Eq.~\ref{W}, multiplying by $t$ and then integrating over $t$:
\begin{equation}
\label{tau}
\int_0^\infty t\frac{\partial^2 W}{\partial t^2}\mathrm{d}t =\nabla \cdot \big[ D(\mathbf{r})\nabla \bar{\tau}(\mathbf{r}) \big] - D(\mathbf{r}) \beta \nabla u(\mathbf{r}) \cdot \nabla \bar{\tau}(\mathbf{r}).
\end{equation}
Integrating by parts the left-hand side of this equation leads to
\begin{align}
\label{tautau}
\int_0^\infty t\frac{\partial^2 W}{\partial t^2}\mathrm{d}t & = \left.t\frac{\partial W}{\partial t} \right|_0^\infty - \int_0^\infty \frac{\partial W}{\partial t}\mathrm{d}t \nonumber \\
& = W(\mathbf{r},0)-W(\mathbf{r},\infty)=-1.
\end{align}
The resulting differential equation for the MFPT is
\begin{equation}
\label{eq:MFPT_1}
\nabla \cdot \big[ D(\mathbf{r})\nabla \bar{\tau}(\mathbf{r}) \big] - D(\mathbf{r}) \beta \nabla u(\mathbf{r}) \cdot \nabla \bar{\tau}(\mathbf{r})=-1.
\end{equation}
Rewriting this equation to the particular case of the spherical geometry, with the volume $V$ representing a sphere of radius $R$, we obtain 
\begin{equation}
\label{eq:MFPT_2}
\frac{1}{r^2}\frac{\mathrm{d}}{\mathrm{d}r} \bigg[ r^2D(r)\frac{\mathrm{d}\bar{\tau}}{\mathrm{d}r}
\bigg] -D(r)\frac{\mathrm{d}(\beta u)}{\mathrm{d}r}\frac{\mathrm{d}\bar{\tau}}{\mathrm{d}r} =-1,
\end{equation}
which can be expressed as
\begin{equation}
\label{eq:MFPT_3}
\frac{\mathrm{d}}{\mathrm{d}r} \bigg[ r^2D(r)e^{-\beta u(r)} \frac{\mathrm{d}\bar{\tau}}{\mathrm{d}r}
\bigg] =-r^2e^{-\beta u(r)}.
\end{equation}

This differential equation can be integrated analytically, using the fact that the MFPT is finite for any point within the volume $V$, and equals zero if the molecule is located at the external surface of radius $R$, $\bar{\tau}(r = R) = 0$. Performing the first integration gives
\begin{equation}
\label{MFPT3}
\frac{\mathrm{d}\bar{\tau}}{\mathrm{d}r} =-\frac{e^{\beta u(r)}}{r^2D(r)}\int_0^r \mathrm{d}r^\prime {r^\prime}^2e^{-\beta u(r^\prime)}.
\end{equation}
Finally, the MFPT for a molecule located at a distance $r = s$ from the center of spherical volume is obtained through the second integration
\begin{equation}
\label{MFPt4}
\bar{\tau}(s)=\int_s^R \mathrm{d}r\frac{e^{\beta u(r)}}{r^2D(r)}\int_0^r {r^\prime}^2e^{-\beta u(r^\prime)}\mathrm{d}r^\prime.
\end{equation}

\begin{acknowledgement}
The authors thank the financial support provided by the Junta de Andaluc\'{\i}a and European Regional Development Fund - Consejer\'{\i}a de Conocimiento, Investigaci\'on y Universidad, Junta de Andaluc\'{\i}a (Projects PY20-00241, A-FQM-90-UGR20) and the Spanish Ministerio de Ciencia e Innovación, Plan Estatal de Investigaci\'on Cient\'{\i}fica, T\'ecnica y de Innovaci\'on (Project PID2022-136540NB-I00).
I.A.B.\ acknowledges María Zambrano Grant funded by MCIN/AEI and NextGenerationEU/PRTR, and the Precompetitive Research Projects Program of the UGR Research Plan (PPJIA2022-46). J.L.M. thanks the Ph.D.\ student fellowship (FPU21/03568) given by Gobierno de España, Ministerio de Universidades. M.K.\ acknowledges financial support from the Slovenian Research Agency ARRS (contract P1-0055). Finally, we thank Prof.\ Gerardo Odriozola (UAM, City of M\'exico) for inspiring discussions and useful comments, and the computational resources and assistance provided by PROTEUS, the supercomputing center of Institute Carlos I for Theoretical and Computational Physics at the University of Granada, Spain.
\end{acknowledgement}


\bibliography{references_doi}


\end{document}